\documentclass[preprint]{aastex}
\usepackage{lscape}
\usepackage{graphicx}
\usepackage{amssymb}
\usepackage{color}

\shorttitle{Rapid SED Variation}
\shortauthors{Ke et al.}

\begin{document}


\title{Rapid Mid-Infrared Variability in Protostellar Disks}


\author{T.T. Ke\altaffilmark{1}, H. Huang\altaffilmark{1}, D.N.C. Lin\altaffilmark{2,3}}


\altaffiltext{1}{School of Physics, Peking University, Beijing
100871, China}
\altaffiltext{2}{Kavli Institute for Astronomy \&
Astrophysics, Peking University, Beijing 100871, China}
\altaffiltext{3}{UCO/Lick Observatory, University of California,
Santa Cruz, CA 95064, USA}


\begin{abstract}
Spectral energy distribution (SED) in protostellar disks is
determined by the disks' internal dissipation and reprocessing of
irradiation from their host stars. Around T Tauri stars, most
mid-infrared (MIR) radiation (in a few to a few ten $\mu$m
wavelength range) emerge from regions around a fraction to a few
AU's.  This region is interesting because it contains both the
habitable zone and the snow line. Recent observations reveal that
SED variations, in the MIR wavelength range.  These variations are
puzzling because they occur on time scale (a few days) which is much
shorter than the dynamical (months to years) time scale at 1AU to a
few AU's. They are probably caused by shadows casted by inner onto
outer disk regions. Interaction between disks and their misaligned
magnetized host stars can lead to warped structure and periodic SED
modulations. Rapid aperiodic SED variations may also be induced by
observed X-ray flares from T Tauri stars. These flares can
significantly modulate the ionization fraction of the gas and the
net charge carried by the grains near the surface of the inner disk.
The newly charged grains may be accelerated by the stellar or disk
magnetic field and adjust their distances from the midplane. Shadows
casted by these grains attenuates the flux of stellar photons
irradiated onto regions at several AU's from the central stars. We
use this model to account for the observed rapid aperiodic SED
variabilities. We suggest regular monitoring of SED variations will
not only provide valuable information on the distribution of the
disk aspect ratio near the habitable zone but also provide a probe
on the interaction between the inner regions of the disk with the
magnetosphere of their host stars.
\end{abstract}


\section{Introduction}
For mid-infrared spectrum in protosteller disks, normally we expect
that the spectral variation is determined by the local disk
dynamics, with typical timescale comparable to the orbital period. Spectrum
around a few ($<10$) $\mu m$ comes from disk irradiation at $0.1\sim1$ AU,
with a typical dynamical timescale as $0.1\sim1$ year. However, a series
of recent observations of T Tauri stars by \textit{Spitzer} revealed
the rapid variability of spectral energy distribution in
protostellar disks with timescales as short as one week. Muzerolle
et al. showed remarkable mid-infrared variability of the
transitional disk LRLL 31 \citep{muz09}, and highlighted the seesaw
feature: "the variability observed at shorter wavelengths is
reversed at longer wavelengths, with a constant pivot point at
$\lambda \sim 8.5\mu m$." The largest flux difference reported is
$60\%$ at short wavelengths and $30\%$ at long wavelengths, and an
impressive change ($20\% - 30\%$) occurred over just one week.
Moreover, the flux variations are fairly constant as a function of
wavelength within intervals $5 - 7 \mu m$ and $11 - 35 \mu m$,
respectively. As a prevailing phenomenon, variability of disk
emission has also been reported in pre-main sequence Herbig Ae stars
\citep{sit08}. Structural changes in the region of the disk near
dust sublimation zone were proposed to account for the variability.
For relatively slow variabilities, Bary et al. (2009) reported
significant variations of $10\mu m$ silicate features on monthly
and yearly timescales. They further pointed out several scenarios
that might produce such variability, including intervening cooler
dust grains, disk shadowing and illumination, turbulent mixing, and
disk winds. Most recently, an extensive study \citep{fla11} has been
carried out for the transitional disk LRLL 31 with multiple epoches
of observations. Based on the observations, authors reasonably put
some constraints on the physical mechanism responsible for the
variability based on their observations: "the variable accretion and
wind are unlikely to be the cause of the variability, nor is a
companion within $\sim 0.4 AU$ perturbing the disk." The most likely
explanation they provided is "either a companion beyond $\sim 0.4
AU$ or a dynamic interface between the stellar magnetic field and
the disk leading to a variable scale height and/or warping of the
inner disk."

Following these puzzling observations, several mechanisms have been
proposed to explain the rapid mid-infrared variability. Flaherty and
Muzerolle considered non-axisymmetric structure variations in the
inner disk \citep{fla10}, such as a warp or a spiral wave, and
suggested the observed variability could be explained by a warped
inner disk with variable scale height. As far as we know, this is
the only paper trying to construct a physical model to explain the
rapid mid-infrared variability at present. Though their model can
generate the pivot point in the SED with a consistent timescale,
their the location of the pivot point is inconsistent with the
observations. Moreover, their model doesn't fit the flux change at
$\lambda > 8\mu m$. Based on the CoRoT light-curves morphology of
$83$ previously known classical T Tauri stars, Alencar at al.
associated the light curve variations with a magnetically controlled
inner disk warp \citep{ale10}. They further pointed out the inner
warp dynamics should result from the interaction between the stellar
magnetic field and the inner disk region. However, they haven't
built up a physical model to consolidate their proposition , and
their explanation concludes a timescale of a few rotational periods,
which usually extends beyond one week, thus inconsistent with the
observations. Espaillat et al. (2011) showed that they can reproduce the
observed variability using irradiated disk models by changing the
height of the inner disk wall by $\sim 20\%$. Their
model builts a reasonable link between the SED variability and the
variation of the inner disk structure.  However, they haven't
provided a physical explanation about how inner disk wall can change
in a short timescale. Based on a self-consistent physical model, we
show in this paper, that observed X-ray flares can account for the
rapid mid-infrared variability in protosteller disks.

Stellar X-ray flares are prevailing among Young Stellar Objects
(YSOs) \citep{fei99, gla05, fav05, get08, ste05}, which may come
from material transfer from disk to the central star, or stellar
magnetic reconnections. A recent COUP survey of the Orion Nebula
Cluster (with Chandra space telescope) indicates that nearly all T
Tauri stars emit X-ray (0.5-8 keV) photons with a nearly constant
quiescent flux in the range of $\sim 10^{23}$J s$^{-1}$. During
COUP's two week observing period, 1-2 flares are commonly observed
around each source\citep{wol05}. These flares are characterized by
power-law energy spectra and up to 1-2 orders of magnitude increase
in their peak X-ray luminosity and they decay exponentially on a
time scale $\sim 10^{3-5}$s. Similar flares (some with larger
amplitudes) were observed with the XEST survey of the Taurus
region\citep{gud07} and the Serpens star-forming
region\citep{gia07}.  They have been modeled with emission from
thermal plasma along cylinderical loops which extend beyond
$R_m$\citep{rea97}. These X-ray flares can greatly influence the
circumstellar disks in various aspects. For example, \citep{wol05}
estimated that the X-ray can ionize protostellar disks out to 1AU at
a rate which is an order of magnitude higher than that estimated for
cosmic rays. In this paper, we focus on the ionization effect of the
X-ray flares.

We develop an analytical framework to show that the observed rapid
X-ray flares from T Tauri stars can significantly modulate the
ionization fraction of the gas and charges carried by the grains
near the surface of  the inner regions of their surrounding disks.
In the proximity of their host stars, the newly charged grains are
accelerated by the stellar and disk field and adjust their distances from
the midplane.  Consequently, the shadow casted by the grains on the
surface of the inner disk regions attenuates in the flux of stellar
photons irradiated at several AU's from the central stars.  We use
this model to account for the variabilities of spectral energy
distribution, at a few micron range, on time scales of a few days.

The rest of this paper is organized as follows. Our primary
objective is to identify and analyze the dominant physical process
rather than to reproduce the SED evolution with the detailed
construction of a new set of disk structure models. In Section 2, we
briefly describe the static disk structure, mainly based on a model
of Dullemond et al. (2001). We analyze X-ray ionization and grain
charge loading process in the inner frontier of the disk in Section
3, and calculate the grain dynamics in Section 4. In Section 5, we
formulate the spectrum variation, and discuss the response-time to
verify the viability of our analysis. In Section 6, we present the
main features of our model through a numerical example, with
comparisons with observations. Finally, in section 7, we conclude
the paper and discuss different scopes in parameter space to further
explore implications of our model.

\section{Disk Model}
In their seminal paper, Chiang and Goldreich (1997) proposed a
hydrostatic radiative equilibrium model for passive disks
surrounding T Tauri stars. The most notable idea they developed
is a "sandwich" model for the vertical structure of the disk, which
consists of an interior layer amidst two surface layers. The
optically thin surface layer is directly heated by the stellar
irradiation, and the interior region is heated by the re-irradiation
from the surface layer. The direct implication of this model is the
flaring configuration of the disk. Dullemond et al. (2001) modified
this model by adding an inner hole due to the evaporation of carbon or
silicon dusts in the disk. In their model, the inner rim of the disk
is puffed-up, because of direct exposition to the stellar flux. As a
consequence, there is region right beyond the inner rim which is in
the shadow. Garaud and Lin (2006) further extended the disk structure by
considering the both internal dissipation and surface irradiation.

In this section, we generally follow the line of Dullemond et al.
(2001) with some modifications to incorporate internal dissipation,
which dominates the energy spectrum in shadowed region. For sake of
simplicity, we write down the main results of their works without
detailed discussions. We consider a T Tauri star with a
circumstellar disk, which is heated by both the irradiation of the
central star and the internal viscous dissipation. Under the
framework of the Minimum-Mass Solar Nebula Model, the surface mass
density can be expressed in the form
$\Sigma=\Sigma_{0}(R/AU)^{-3/2}$, with $\Sigma_{0}=10^5kg/m^2$. The
disk is divided into several regions: the puffed-up rim which is the
inner edge of the disk locating in the sublimation line of
silicates; the shadowed region which lacks the direct irradiation of
the central star by the rim sheltering; the flaring part including
the interior and surface layers. The intermediate shadowed region is
heated by viscous dissipation whereas the outer part is mostly
heated by stellar irradiation.  For protostellar disks with a solar
composition and accretion rate ${\dot M}_d$ around solar mass host
stars, the transitional radius between these two regions is
$r_{tran} \sim 1 ({\dot M}_d / 3 \times 10^{-9} M_\odot
y^{-1})^{0.8}$ \citep{gar06}.

In most parts of the disk, small ($\mu m$-size) grains are thought to
be homogenously mixed with the gas, with the mass density percentage
$Z=\frac{\rho_{grain}}{\rho_{gas}}=1\%$. We take grains to be
spheres with the average radius $r_{d}$ ranging from $0.1\mu m$ to
$1\mu m$, and the density $\rho_{d}\sim 3\times10^{3}kg/m^{3}$. The
emissivity and correspondingly opacity possess two Gaussian peaks
located at $10\mu m$ and $18\mu m$ due to silicate resonance. When
dealing with the heat equilibrium equations, we omit the
contribution of silicate resonance, and use $\varepsilon=T/T_{*}$
and $\kappa=\kappa_{v}T/T_{*}$ instead.

Optical height $H$ and temperature in the interior layer $T_i$ at
radius $R$ can be expressed as
\begin{eqnarray}
H&=&\left(\frac{2}{7}\right)^{\frac{1}{7}}\chi^{\frac{8}{7}}
\left(\frac{\psi_{s}}{2\psi_{i}}\right)^{\frac{1}{7}}
\left(\frac{T_{*}}{T_{0}}\right)^{\frac{4}{7}}
\left(\frac{R}{R_{*}}\right)^{\frac{2}{7}}R,\\
T_{i}&=&\left(\frac{2}{7}\right)^{\frac{1}{7}}\chi^{\frac{2}{7}}
\left(\frac{\psi_{s}}{2\psi_{i}}\right)^{\frac{2}{7}}
\left(\frac{T_{*}}{T_{0}}\right)^{\frac{1}{7}}
\left(\frac{R}{R_{*}}\right)^{-\frac{3}{7}}T_{*}.
\end{eqnarray}
The dimensionless constant $\chi$ varies between 2 and 6 in the
region discussed, which can be determined with an appropriate
definition of the optical height \citep{gar06}. Functions $\psi_{i}$
and $\psi_{s}$ depend on the temperature in the interior layer and
surface layer respectively, accounting for the possibility that the
disk layer is not fully optically thick to its own emission
\citep{dul01}. The symbols $T_{*}$ and $R_{*}$ represent,
respectively, the surface temperature and radius of the central
star. Constant $T_{0}=\frac{GM_{*} m_p}{k_B R_{*}}$ in unit of
temperature, depends on stellar mass $M_{*}$ and radius $R_{*}$, and
mass of hydrogen atom $m_p$.

Temperature and surface mass density for surface layer can be
expressed as
\begin{eqnarray}
T_{s}&=&\left(\frac{R_{*}}{2R}\right)^{\frac{2}{5}}T_{*}, \\
\Delta\Sigma
&=&\frac{\alpha(\frac{R_{*}}{R})^{2}T_{*}^{4}}{2\kappa_{p}(T_{s})T_{s}^{4}},
\end{eqnarray}
where $\alpha=\frac{2H}{7R}$ is the grazing angle of the irradiation
from the central star to the surface of the flaring disk, and
$\kappa_{p}(T_s)$ is the Planck mean opacity for surface layer.

Assuming the majority of dust grains in the disk as silicates, with
evaporation temperature $T_{evp}\simeq1500K$, the disk will be
truncated in a special radius $R_{rim}$, inside which, the grains
are in the gaseous state mixed with gas of light elements and are
optically thin to the stellar irradiation. In the case that
self-irradiation cannot be ignored, the location and optical height
of the rim can be jointly determined by following equations:
\begin{eqnarray}
R_{rim}&=&R_{*}\left(\frac{T_*}{T_{evp}}\right)^{2}
\left(1+\frac{H_{rim}}{R_{rim}}\right)^{\frac{1}{2}},\label{eq:tempblnc}\\
H_{rim}&=&\chi_{rim}\left(\frac{T_{evp}}{T_{0}}\right)^{\frac{1}{2}}
\left(\frac{R_{rim}}{R_{*}}\right)^{\frac{1}{2}}R_{rim},\label{eq:hrim}
\end{eqnarray}
where the dimensionless constant $\chi_{rim}$ is different from that
of the interior layer of the flaring disk because of the different
grain density distribution and disk structures. From the discussion
in Garaud and Lin (2006), one can expect that $\chi_{rim}$
will be some bigger than $\chi$, due to the optically thin inner
hole.

The above expression for $R_{rim}$ is different from the truncation
radius for the disk gas.  Many T Tauri stars have strong (kG) fields
on their surface\citep{joh04, joh07}.  These fields dominate the gas
flow and truncation protostellar disks out to a distance
\begin{equation}
R_m = \beta \mu^{4/7} {\dot M}_d ^{-2/7} (2 G M_\ast) ^{-1/7}
\end{equation}
where $\mu$ is the dipole magnetic moment, ${\dot M}_d$ is the
accretion rate in the disk, and dimensionless parameter $\beta$
ranges from $0.5$ to $1$ \citep{kon91}. Since grains are the
primarily opacity sources, the dust destruction radius is more
relevant for radiation transfer and the location of $R_{rim}$
provided it is larger than $R_m$.

In most regions of the disk, especially at large radii, energy
produced by the viscous dissipation is relatively small compared
with stellar irradiation.  For computational convenience, we consider
the contribution of the frictional heat only in the shadowed region,
stretching from the inner rim to the outer boundary $R_{fl}$. The
following result further justifies this simplification. By equating
$\frac{H_{rim}}{R_{rim}}$ with $\frac{H_{fl}}{R_{fl}}$ we obtain the
locus of boundary at
\begin{equation}
R_{fl}=\left(\frac{2}{7}\right)^{-\frac{1}{2}}\chi^{-4}
\left(\frac{\psi_{s}}{2\psi_{i}}\right)^{-\frac{1}{2}}
\left(\frac{T_{*}}{T_{0}}\right)^{-2}
\left(\frac{H_{rim}}{R_{rim}}\right)^{\frac{7}{2}}R_{*}.\label{equ:Rfl}
\end{equation}
Following the framework developed by Ruden and Lin \citep{rud86} on
the basis of the $\alpha-$disk model \citep{sha73}, we obtain the
effective temperature $T_{e}$ in the shadowed region as
\begin{equation}
T_{e}= \left(\frac{3\sum_{0}\alpha_{s}k_B}{4\sigma
m_p}\right)^{\frac{1}{3}}
\left(\frac{GM_{*}}{AU^{3}}\right)^{\frac{1}{6}}
\left(\frac{R}{AU}\right)^{-1},
\end{equation}
where the dimensionless efficiency factor $\alpha_s\simeq 0.001$.

\section{X-Ray Ionization and Grain Charging}
Absorbed by grains and gas in the disk, the stellar X-ray can only
penetrate a certain distance into the disk just like the visual
photons do.  Except for the directly exposed inner rim and the
flaring surface layer, the structure of other regions in the disk
is not significantly influenced by the X-ray irradiation.

\subsection{Ionization state of the gas}

The rate of change in ionization fraction of the gas is
\begin{equation}
\frac{\partial x}{\partial t} = (1-x) \left[ \frac {\Gamma} {k_B
T_\ast} + e(T) nx \right] -\left[\alpha_B (T) n x^2 + \frac{Z
\rho}{m_{d}} S_ev_{e}\sigma_{e}(Q)x\right] \label{eq:ionizex}
\end{equation}
where $\Gamma$ is the ionizing energy input rate in J s$^{-1}$
H-atom$^{-1}$, $T_\ast$ is the averaged source temperature,
$\alpha_B (T) = 3 \times 10^{-16} T^{-3/4}$ photons m$^3$ s$^{-1}$
is the case B recombination rate \citep{ost06}, $e(T)=6 \times
10^{-17} T^{1/2}e^{-E_0/k_BT}$ m$^3$ s$^{-1}$ is collisional
ionization rate \citep{dra11}, and $n=\frac{\rho}{m_p}$ is the total
number density of hydrogen. In the above equation, the last term
takes into account the recombination of protons and electrons after
they collide with the grains. $S_e$, $v_e$, and $\sigma_e$ are the
absorption rate, velocity dispersion, and grains' cross section for
electron capture which may be a function of the grains' net charge
(see subsections below).

In the above equation, only photons that are more energetic than the
ground state energy of the hydrogen atom, $E_{0}=13.6eV$ contribute
to the flux of ionizing photons. The observed energy
distribution\citep{hud91,cro93,wol05} in the X-ray range suggests
that T Tauri stars emit photon flux (ie photons per $m^2$ per
second)
\begin{equation}
\frac{d{\dot N}_\gamma (E)}{dE_\gamma} = \frac{{\dot N}_0}{E_0} \left(
\frac{E}{E_0} \right)^{-1.7}
\label{eq:xsed}
\end{equation}
where the normalization factor ${\dot N}_0$ can be obtained from the
total X-ray luminosity $L_X$ in the energy range between $E_0$ and
$E_{max}$ (which is 8kev for the COUP sources),
\begin{equation}
\frac{L_X}{4 \pi R_{rim}^2} = \int_{E_0} ^{E_{max}} E \frac{d {\dot
N}_\gamma}{d E_\gamma} dE = \frac{{\dot N}_0 E_0}{0.3}\left[ \Big(
\frac{E_{max}} {E_0} \Big)^{0.3}-1\right].\label{eq:lx0}
\end{equation}

In equation (\ref{eq:ionizex}), the heating and ionization input rate
\begin{equation}
\Gamma= \int_{E_0}^{E_{max}} E\sigma_X (E) \frac{d{\dot N}_\gamma
(E)}{d E} dE
\end{equation}
\begin{equation}
\frac{\Gamma}{k_B T_{\ast}} = \int_{E_0}^{E_{max}} \sigma_X (E)
\frac{d{\dot N}_\gamma (E)}{d E} dE \label{eq:gammakt}
\end{equation}
where the X-ray photoionization cross section
$\sigma_{X}(E)=\widetilde{\sigma}(E/E_0)^{-2.5}$, for photons'
energy $E$, and $\widetilde{\sigma}=10^{-21}m^2$\citep{gla97}. From
equations (\ref{eq:xsed}), (\ref{eq:lx0}), and (\ref{eq:gammakt}),
we find
\begin{equation}
\frac{\Gamma}{k T_\ast} = \frac{{\dot N}_0 \widetilde{\sigma}}
{3.2}\left[ 1-\Big( \frac{E_{max}} {E_0} \Big)^{-3.2}\right] \simeq
\frac{7.5 \times 10^{-3} L_X \widetilde{\sigma}} {R_{rim}^2 E_0}
\left(\frac{E_{\max}}{E_0} \right)^{-0.3}. \label{gammakt}
\end{equation}
This expression can be used to determine the changing rate of
ionization fraction in equation (\ref{eq:ionizex}).

The internal thermal energy ($E_g$) equation of the gas can be
written as \citep{lin92}
\begin{equation}
m_p \left[ \frac{d E_g}{d t} + p \frac{d (1/\rho)}{d t} \right]=
\Gamma (1-x) - n \Lambda (x, T) - n \left[ \frac{3}{2} x^2 k_B T
\alpha_B (T) + x (1-x) e(T) E_0 \right] \label{eq:thermale}
\end{equation}
where $p = \rho k T (1+x) / m_H$ is the pressure. At temperatures
below $10^4$ K, the cooling function $\Lambda(x, T)$ is given by
Dalgarno \& McCray (1972) for optically thin line emission by metal
ions with a solar abundance. Similar to equation (\ref{eq:gammakt}),
we replace $\Gamma$ with
\begin{equation}
\Gamma =\frac{{\dot N}_0 \widetilde{\sigma}} {2.2}\left[ 1-\Big(
\frac{E_{max}} {E_0} \Big)^{-2.2}\right] \simeq \frac { 10^{-2} L_x
\widetilde{\sigma}} {R_{rim}^2} \left(\frac{E_{\max}}{E_0}
\right)^{-0.3}.
\end{equation}
In our model, the ionization fraction and the gas temperature of the
gas are determined by the intensity of the ionizing photons emitted
near the central star.  During quiescence when the X-ray luminosity
is less than $\sim 10^{23}$J s$^{-1}$, the surface of the disk
region at $\sim 0.1$ AU retains its low gas temperature and
ionization fraction.  In this state, contribution from the
collisional ionization rate $e(T)$ is negligible because free
electrons and protons are more likely to collide with grains than
directly with each other to recombine. With their relatively large
cross section, grains easily can capture ionized electrons and
protons, and become negatively charged.

However, during some outbursts when the flux of X-ray increases to
$\sim 10^{25}$ J s$^{-1}$. Nevertheless, $L_X$ is generally small
compared with the stellar luminosity at longer wavelengths such that
$T_i$, $T_s$, and $T_e$ are not affected by them.  However, X-ray
outbursts can significantly increase ionization fraction and the gas
temperature, perhaps trigger the onset of thermal instability
\citep{mur92} in the tenuous upper layers the disk at $\sim 0.1$ AU.
In this limit, collisions between the electrons and protons are
important in the determination of the ionization fraction.
Electrons' thermal energy is also elevated.

Now let's start estimating the equilibrium temperature and
ionization fraction of the disk gas during quiescent state
and X-ray flares.  We show that equilibrium can indeed be reached within
a relatively short time compared with the daily timescale of X-ray
flare duration. By taking the right hand side of both equation
(\ref{eq:ionizex}) and (\ref{eq:thermale}) to be zero, we can solve
the equilibrium temperature and ionization fraction as $T_q \simeq
472 K$ and $x_q \simeq 0.038$ in the quiescent state when the X ray
luminosity $L_X \simeq 5\times 10^{22}J/s$. Note that the value
of $T_q$ is small compared with $T_i$ produced by both viscous
dissipation and reprocessed stellar photons in the visual wavelength
range.  Therefore the X-ray photons do not significantly modify the
disk structure and ionization fraction during the quiescence.

However,  the equilibrium
temperature and ionization fraction as $T_f \simeq 9700 K$ and $x_f
\simeq 0.79$ during the X-ray flares when the X ray luminosity
increases to $L_X \simeq 1\times 10^{25}J/s$. In this phase,
$T_f > T_i$ and photoionization by the X-ray photons determines
the thermal structure of the disk surface layer.  In the above
estimation, we are looking at gas/plasma with number density $n
\simeq 10^{16}m^{-3}$ at $6$ scale heights from the midplane
($\chi_{rim}=6$) in the frontier of inner rim located around $0.1AU$
from the central star (refer to the numerical section). The optical
radial thickness of the inner rim is around $0.04AU$ at $6$ scale
heights, which is comparable with $R_{rim}\sim 0.1 AU$ and
$H_{rim}\sim 0.02 AU$. Consistent with our previous intuition, both
temperature and ionization get elevated during X-ray flares.

For dynamics of gaseous ionization by X-ray flares, we find the
ionization fraction goes as $\big(1-e^{-t/\tau_i}\big)$ if we
neglect the collisional ionization and direct recombination, with
the response time
\begin{equation}
\tau_I=\Big(\frac{\Gamma}{k_B T_*} + \frac{Z \rho_{rim}}{m_d}S_e v_e
\sigma_e\Big)^{-1}\sim 10^{2}s
\end{equation}
which is much shorter than one day. By taking into account the
collisional ionization and direct recombination, the ionization
should proceed even faster. For heating and cooling dynamics, we
estimate the response time respectively from the equation
(\ref{eq:thermale}) is of the order
\begin{eqnarray}
\tau_H &=& \frac{k_B(T_f-T_q)}{\Gamma} \sim 1s\\
\tau_C &=& \frac{k_B(T_f-T_q)}{n \Lambda(T_f,x_f)} \sim 10s
\end{eqnarray}
which are again much shorter than the daily X-ray flare timescale.
Finally we show that blackbody irradiation for grains is so
effective that the conductive heat transfer from overheated ambient
gas to grains is negligible. In fact, during X-ray flares, grains
are heated not only by stellar flux, but also by the collisions with
overheated gas. Thus instead of equation (\ref{eq:tempblnc}), we
have
\begin{equation}
\sigma T_{grain}^4 =\sigma \Big(\frac{R_{*}}{R_{rim}}\Big)^2 T_*^4
\Big(1 + \frac{H_{rim}}{R_{rim}}\Big)+6 n v_H k_B (T_f - T_{grain})
\end{equation}
where $\sigma$ is the Stefan-Boltzmann constant, and average
velocity of hydrogen atoms is approximated as $v_H=\sqrt{\frac{3k_B
T_f}{m_p}}$. For ambient gas/plasma temperature during X-ray flares
$T_f\simeq 9700K$, the equilibrium temperature for grains
$T_{grain}=1500K$ which is comparable to the equilibrium temperature
$T_{evp}=1500K$ in the quiescent state. Thus conductive heat
transfer from ambient gas is entirely negligible compared with the
stellar irradiation in determining temperature of grains in the
inner rim. Last but not least, after the X-ray flare, the rise in
the gas temperature implies that the disk scale height will increase
by a factor of two on the local dynamical time scale (probably a few
days), which will further lower the local gas density. As a result,
the equilibrium temperature should be higher, and more charge will
be loaded on grains. Thus grains experience a stronger magnetic
force and a smaller gas drag, but at the same time a outward gas
push. It's complicated to judge which effect dominates, and we
suspect all to be insignificant. So in this paper, we omit the
influence of the change of gas scale height right after X-ray
flares.

\subsection{Charge-loading process}
We now consider the charge carried by the grains.
X-ray flares charge grains mainly through two mechanism: collisional
charging and photoelectric emission \citep{dra11}. Photoelectric
emission happens when a X-ray photon excites an energetic electron
out of the grain surface. However the effect of photoelectric
emission is comparable with collisional charging only when the
photoelectric yield is of order unity. Hindered by multiple
complications, photoelectric yield is estimated around $0.1$ for the
grain size and photon energy we considered \citep{wei06}, thus
negligible. We omit the photoelectric emission and focus on the
collisional charging effect below.

In the framework of a certain grain with charge $Q$, the cross
sections for electrons and protons to collide with this grain are
\citep{ume83}:
\begin{eqnarray}
\sigma_{e}&=&\pi r_{d}^{2} e^{\frac{1}{4\pi\varepsilon_{0}}\frac{eQ}
{r_{d}E_{e}}}\\
\sigma_{p}&=&\pi
r_{d}^{2}\Big(1-\frac{1}{4\pi\varepsilon_{0}}\frac{eQ}{r_{d}E_{p}}\Big),
\end{eqnarray}
where $e$ is the elementary electric charge, $\varepsilon_0$ is
vacuum permittivity, $E_{e}=\frac{1}{2}m_{e}v_{e}^{2}$ and
$E_{p}=\frac{1}{2} m_{p}v_{p}^{2}$ are the kinetic energy of protons
and electrons respectively. As shown above, the temperature
equilibrium can be achieved in a relatively short time, so that the
energy should be equally partitioned between electrons and protons
$E_{e}\simeq E_{p}=\frac{3}{2}k_B T_f$.

By introducing $S(e)$ and $S_{p}$ as the absorption rate for the
electrons and protons after their collision with grains.  We write
the microscopic charge balance equation for grains:
\begin{equation}
S_{e}
n_{e}v_{e}\sigma_{e}=S_{p}n_{p}v_{p}\sigma_{p},\label{equ:balance}
\end{equation}
where $n_e$ and $n_p$ are number density of protons and electrons
respectively. By defining dimensionless parameter
\begin{equation}
\theta =-\frac{1}{4\pi\varepsilon_{0}}\frac{Q e }{r_{d} E_{e}},
\end{equation}
we can rewrite (\ref{equ:balance}) as
\begin{equation}
\frac{e^{-\theta}}{1+\theta}=\frac{n_{p}}{n_{e}}
\frac{S_{p}}{S_{e}}\Big(\frac{m_{e}}{m_{p}}\Big)^{\frac{1}{2}}.\label{equ:chargeloading}
\end{equation}

High speed electrons tend to have a lower absorption rate than that
for protons, and we estimate $0.01\lesssim\frac{S_e}{S_p}\leq1$
\citep{ume83}. Moreover as inferred from ionization fraction $x_f
\simeq 0.79$, during X-ray flares, the electrons and protons are so
abundant that the charge on grains consist of only a tiny portion of
the entire plasma, thus we have $n_e=n_p$. With these restrictions,
we find from equation (\ref{equ:chargeloading}) $2.5 \lesssim \theta
\lesssim6.4$, which is entirely determined by the ratio of
absorption rates for electrons and protons. We adopt $\theta=5$
(with $S_e=0.05$) in the following analysis. Conclusively we know in
the equilibrium state, grains are negatively charged to
\begin{equation}
Q_0=-5\frac{4\pi\varepsilon_{0}r_{d}E_{e}}{e}\label{equ:Q0}
\end{equation}
which is proportional to the radius of the grain $r_d$ and the
electrons' energy $E_e$, but independent of the ionization fraction
and electron density. During X-ray flares, plasma temperature in the
inner rim increases dramatically, leading to a increase in $E_e$
from $\frac{3}{2}k_B T_q$ to $\frac{3}{2}k_B T_f$, and eventually a
propositional increase in $Q_0$ by $\frac{T_f}{T_q}\sim 20$ times.

Dynamics of charge loading processes can be determined by the
following differential equation
\begin{equation}
\frac{dQ}{dt}=-eS_e v_{e}\sigma_{e}(Q)n_e+eS_p v_{p}\sigma_{p}(Q)
n_{p}\label{equ:dQ}
\end{equation}
We can determine the response time for charge loading on grains to
be
\begin{equation}
\tau_Q\sim\frac{\varepsilon_{0}E_0}{e^2 n_e S_e v_e r_d}\sim10^{-7}s
\end{equation}
which is very swift, thus it takes nearly no time for the grains to
get charged.

In conclusion, we show in this section that X-ray flares will not
only significantly ionize gas in the inner rim of the circumstellar
disk, but also elevate its temperature by thousands of Kelvin
higher. Both ionization and temperature elevation will accomplish in
a much short timescale compared with X-ray flare duration. The
overheated gas and plasma are shown to have negligible effect on
equilibrium temperature of the grains, but will significantly
increase charge on the grains in a relatively short time. In next
section, we will show that the newly charged grains will undergo a
collective migration driven by stellar magnetic field, thus modulate
the height of inner rim of the disk.

\section{Grain Dynamics}
When the temperature in the inner rim is high enough ($\sim 10^3K$)
for collisional ionization of metal atoms, magnetic field can be
coupled effectively with the gas in the inner region of a
circumstellar accretion disk. However despite of voluminous studies
on the interaction of stellar magnetic field and the circumstellar
disk \citep{pri72,gho79a,gho79b,aly80,kon91,liv92,shu94,hay96},
there is little consensus on how well the plasma couple with the
field lines. Some outstanding issues include "whether
the disk excludes the stellar magnetic field by diamagnetic currents
or the field can penetrate and thread a large fraction of the disk,
whether the threaded field remains closed (connecting the star and
the disk) or becomes open by differential shearing \citep{lai99}."
There has also been a recent trend on 3D-modeling magnetic stellar
field \citep{gre07,jar07}, which is very complicated and beyond the
scope of our analytical framework. In this section, we first develop
a parametric framework for the magnetic field, taking into account
the interactions between the stellar field and the circumstellar
disk; and then discuss the charged grain dynamics driven by the
magnetic field.

As mentioned above, the circumstellar disk would be disrupted at the
magenetospheric boundary $R_{m}$ where the magnetic stress is large
enough to remove the excess angular momentum of the nearly Keplerian
flow over a narrow transition zone and channel the plasma onto the
stellar polar caps. The dynamic consequence of disk accretion onto
the central star is to synchronize the stellar spin with the disk
rotation. How much time it takes to reach an equilibrium state
depends on the coupling intensity of field line and the disk. The
corotational radius $R_{co}\sim R_{m}$, and Konigl adopted
$R_{co}=2R_{m}$ \citep{kon91}, and in fact the ratio can be even
smaller. Within the range from $R_{co}$ to $R_{m}$ the disk angular
velocity departs significantly from the Keplerian value, and
corotates with the central star; while beyond $R_{co}$, the angular
velocity is Keplerian \citep{gho79a,gho79b}. For a typical T Tauri
stars, the spin period is around $8$ days, so that $R_{co}\sim
0.07AU$, which as shown later, is smaller than $R_{rim}$ the
sublimation frontier of the disk. In our parametric framework, we
denote
\begin{equation}
R_{co}=\epsilon R_{rim},\label{eq:epsilon}
\end{equation}
where the dimensionless parameter $\epsilon$ ranges from $0.3$ to
less than $1$. In order for the stellar magnetic field to induce
a spin equilibrium (in which there is little net angular momentum
flux between the stellar spin and the disk), it must be well coupled
to the disk beyond the corotation radius (Ju et al in preparation).

We assume the stellar field can influence the surface layer of the
disk near $R_{rim}$.  We consider the case that stellar magnetic
dipole, stellar spin and the disk angular momentum are aligned with
each other. In the disk mid-plane, the magnetic field in the
cylindrical coordinate can be expressed as:
\begin{eqnarray}
B_z&=&-B_s\frac{R_*^3}{R^3},\\
B_{\varphi}&=&\mp\zeta B_s\frac{R_*^3}{R^3},\label{eq:zeta}\\
B_{R}&=&0,
\end{eqnarray}
where $B_s\sim10^3Gs$ represents the magnetic strength at the
equator at the surface of the central star. The negative sign for
the longitudinal component $B_z$ comes from our consideration that
the magnetic moment is parallel to the stellar spin. We leave the
alternative anti-parallel case in the below discussion section. The
azimuth component $B_{\varphi}$ comes from twisting of $B_z$, and
the upper (lower) sign corresponds to the value at the upper (lower)
disk surface. The quantity $\zeta$ specifies the azimuthal pitch of
the field line. In general we expect $\zeta\lesssim 1$, but its
actual value or form depends on details of the dissipative processes
involved in the disk-magnetic field interactions. If the stellar
magnetic field threads the disk in a closed configuration, we expect
$\zeta\propto(\omega_s-\omega_d)$ \citep{liv92}, where $\omega_s$
and $\omega_d$ are angular momentum for central star and disk
respectively . However, it has been argued that the differential
shearing and the plasma flowing from the disk into the overlying
magnetosphere will blow the field lines open and maintain them in an
open configuration, in which case we expect $\zeta$ to be positive
and of order unity \citep{lai99,hay96}. The radial component
$B_{R}$ comes from the inward accretion flow with velocity $v_R\sim
\nu/R$ which is much smaller than the azimuthal disk velocity, so we
adopt $B_{R}=0$ in the analysis. Because
$\frac{H_{rim}}{R_{rim}}\ll1$ as shown in the following numerical
example, when we discuss the grain dynamics, we can always
approximate magnetic field  as
\begin{equation}
\mathbf{B}=\mathbf{B}_{D}\mp\zeta
B_s\frac{R_*^3}{R^3}\mathbf{\hat{\varphi}}
\end{equation}
in the proximity of disk mid-plane, where $\mathbf{B}_{D}$ is the
magnetic field by unperturbed stellar dipole.

After grains in the inner rim get charged by X-ray ionization, they
move upward (downward in antiparallel case) collectively driven by
the magnetic field. As a consequence, both the shape and the
location of the inner rim changes over time, with the scale of
movement as the Lorentz radius $R_{L}=\frac{u m_{d}}{B_D Q_{0}}$,
and the timescale as the Lorentz period $T_{L}=\frac{2 \pi
m_{d}}{B_D Q_{0}}$. We focus on the dynamics of grains which
originally settle around the upper/lower brim of the inner rim, and
try to infer the variability of $H_{rim}$ based on these grains'
migration.

The equation of motion for a certain grain in the rotating framework
of central star can be formulated as
\begin{equation}
m_{d}\dot{\mathbf{v}}_r=-m_{d}\frac{GM_{*}\mathbf{R}}{r^{3}}+Q_{0}
\mathbf{v_r}\times\mathbf{B}-\nu |\mathbf{v}-\mathbf{v_K}|
(\mathbf{v}-\mathbf{v_K})-2m_d
\mathbf{\omega_s}\times\mathbf{v_r}-m_d
\mathbf{\omega_s}\times(\mathbf{\omega_s}\times \mathbf{r}),
\end{equation}
where $\mathbf{\omega_s}$ is the spin angular velocity for central
star, $\mathbf{v_s}=\mathbf{\omega_s}\times \mathbf{R}$ is the
corotation velocity with the central star,
$\mathbf{v_r}=\mathbf{v}-\mathbf{v_s}$ is grain's velocity in the
rotating framework, and $\mathbf{v_K}$ is the Keplerian velocity.
The last two terms are  Coriolis and centrifugal force in the
rotating framework. We only preserve the radial component of
gravity, because the vertical component has been balanced with the
gas turbulence, and that's the reason why the grain can stay around
the brim initially. We neglect collisions between grains.
Gas drag coefficient $\nu=\frac{1}{8}\pi r_d^2
\rho$ is calculated based on the fact that the velocity of the grain
is much larger than the thermal velocity of the gas, so the gas can
be treated as still relative to the fast grains. Hermite scheme
\citep{aar03} can be adopted to numerically solve the differential
equation, with constant time steps much smaller compared with the
Lorentz period. Notice that the Lorentz force decreases as
$R^{-\frac{7}{2}}$, while the gravity decreases as $R^{-2}$ . For
the grains in the surface layer of the flaring disk around
$R\gtrsim3AU$, the relative strength of the Lorentz force with
respect to gravity becomes $\sim 10^{-3}$ of the value at the inner
rim. Thus we can omit the magnetically driven dynamics of grains in
the surface layer of the flaring region. After we get the grain
dynamics, we approximate the variability of $H_{rim}$ as the
location variations of grains in the brim.

\section{Spectrum Variability}
The collective outward migration of grains (in the parallel case)
puffs up inner rim of the disk. Consequentially, the irradiation
flux from the inner rim increases; while larger region in the
flaring disk becomes shadowed by the rim. We track the flux
variability of both inner rim and shadowed region, by tracking the
variability of the height of inner rim $H_{rim}$. As shown below,
the blackbody irradiation for individual grains is so effective that
it takes almost no time for the newly shadowed (exposed) grains to
reach a new equilibrium temperature. So we can determine the
boundary $R_{fl}$ dividing the shadowed region and the exposed
region by enabling geometric relationship instantly, the same as the
static case (\ref{equ:Rfl}):
\begin{equation}
\frac{H(R_{fl})}{R_{fl}}=\frac{H(R_{rim})}{R_{rim}}
\end{equation}
Once the variability of disk structure determined, computation of
the spectra before and after onset of X-ray flares is pretty
standard \citep{dul01}.

Finally we justify the temperature response time for newly
exposed/shadowed grains in the flaring part. For newly exposed
grains in the surface layer, the temperature dynamics follows
\begin{equation}
\pi r_{d}^{2}\sigma T_{*}^{4}(\frac{R_{*}}{R})^{2}-4\pi
r_{d}^{2}\varepsilon\sigma T_{s}^{4}=\frac{4}{3}\pi
r_{d}^{3}\rho_{d}c_{d}\frac{dT_{s}}{dt}.
\end{equation}
For grains around $3AU$, it's not hard to figure out the temperature
response time is around $10^{-3}$ day, which is much shorter than
the spectrum variability with daily timescales as we are interested
in. Similarly, the newly shadowed grains in the surface layer
undergoes a rapid drop of temperature with timescales around $1s$,
much shorter than the daily timescale. As the irradiation from the
surface layer can only penetrate into a distance in the interior
layer, the newly exposed surface layer will heat a thin top interior
layer quickly by the irradiation, without changing the temperature
of the inner interior layer much. The surface mass density of
quickly heated top interior layer $\Delta\Sigma_{i}=\rho_{i}
l=\frac{\tau_{v}}{\kappa_{v}}\sim\frac{1}{\kappa_{v}}$, then the
temperature dynamics follows
\begin{equation}
\psi_{s}\alpha\sigma T_{*}^{4}(\frac{R_{*}}{R})^{2}-2\psi_{i}\sigma
T_{i}^4=\frac{1}{\kappa_{v}}(\frac{C_{v}}{\mu_{H}}+Zc_{d})\frac{dT_{i}}{dt},
\end{equation}
where $C_{v}=\frac{3}{2}R_{gas}$ is the gas specific heat per mol.
We estimate the temperature response timescale for grains in the
interior layer around $3AU$ around $10^{2}$ days, which is much
longer than the daily timescale we are interested in. In the above
discussion, we omit the effect of the vertically convective heat
transference. Since it has the time scale comparable to the local
dynamics, the effect of convective heat transference should works at
at timescale much longer than several days. Conclusively, the newly
exposed/shadowed grains in the surface layer change instantly; while
grains in the interior layer change little in the daily timescale.

\section{Numerical Studies}
In this section, we apply our model to an numerical example to
generate predictions and compare with observations. We take the
central star as a spherical blackbody with $M_{*}=1.8M_{\bigodot}$
\citep{muz09}, $R_{*}=3R_{\bigodot}$, $T_{*}=4000K$. The outer
cutoff radius of the disk $R_{out}=2.3\times10^{2}AU$ \citep{chi97}.
The radius of grains $r_{d}=0.5\mu m$. The value of $\chi$ takes
different in different regions, where $\chi=4$ for the flaring disk
and $\chi_{rim}=6$ for the inner rim due to a higher gas density
\citep{dul01, dul10, gar06}. The stellar magnetic field at surface
$B_s=1000Gs$, the azimuthal component coefficient in equation
(\ref{eq:zeta}) $\zeta=0.5$ and dimensionless coefficient in
equation (\ref{eq:epsilon}) $\epsilon=0.5$.

\begin{figure}[htbp]
\begin{center}
\includegraphics[width=14cm,height=8cm]{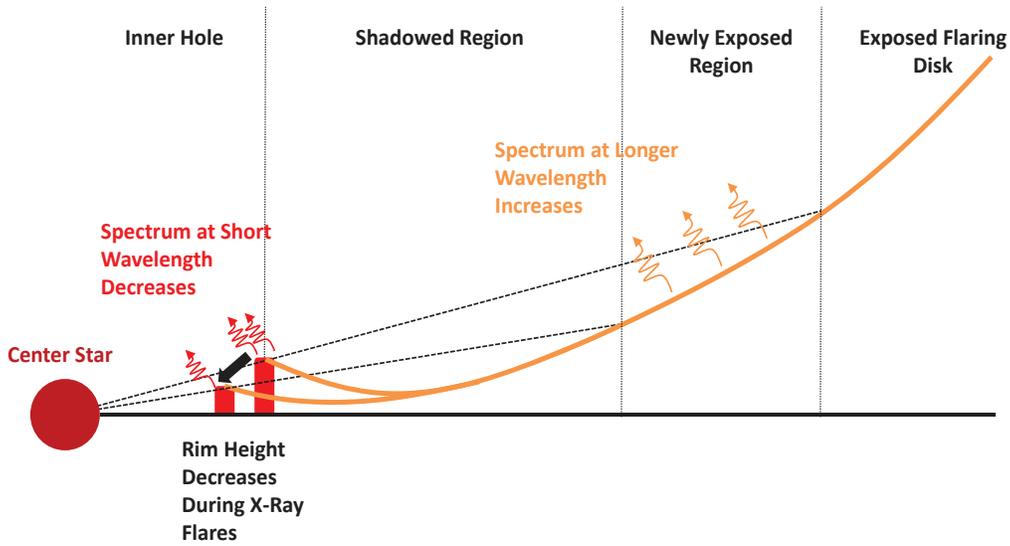}
\caption{A sketch illustrating the spectral variability resulting
from disk structural variability during X-ray
flares.}\label{fig:sketch}
\end{center}
\end{figure}

\begin{figure}[htbp]
\begin{center}
\includegraphics[width=14cm,height=8cm]{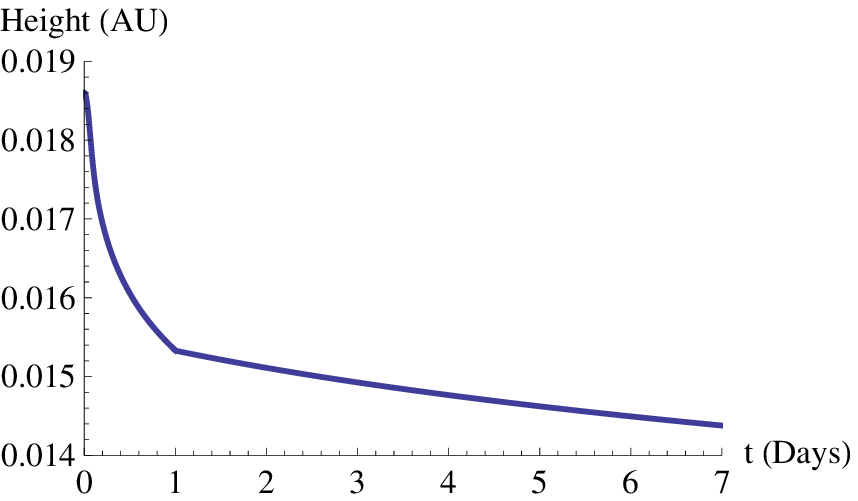}
\caption{Height of inner rim decreases over time, in which case
X-ray flare occurs at time $t=0$ day and ends at time $t=1$
day.}\label{fig:Ht}
\end{center}
\end{figure}

\begin{figure}[htbp]
\begin{center}
\includegraphics[width=14cm,height=8cm]{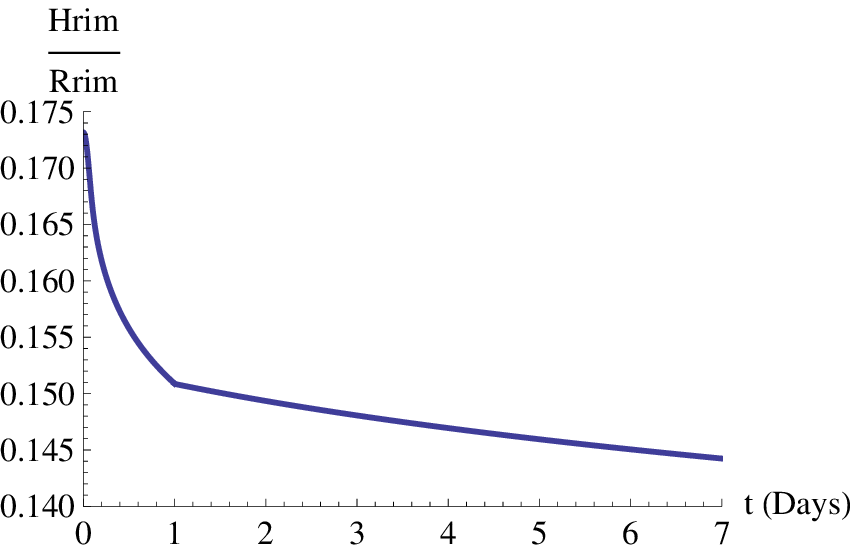}
\caption{Ratio of height over radius of the inner rim decreases over
time, in which case X-ray flare occurs at time $t=0$ day and ends at
time $t=1$ day.}\label{fig:HRt}
\end{center}
\end{figure}

\begin{figure}[htbp]
\begin{center}
\includegraphics[width=11cm,height=10cm]{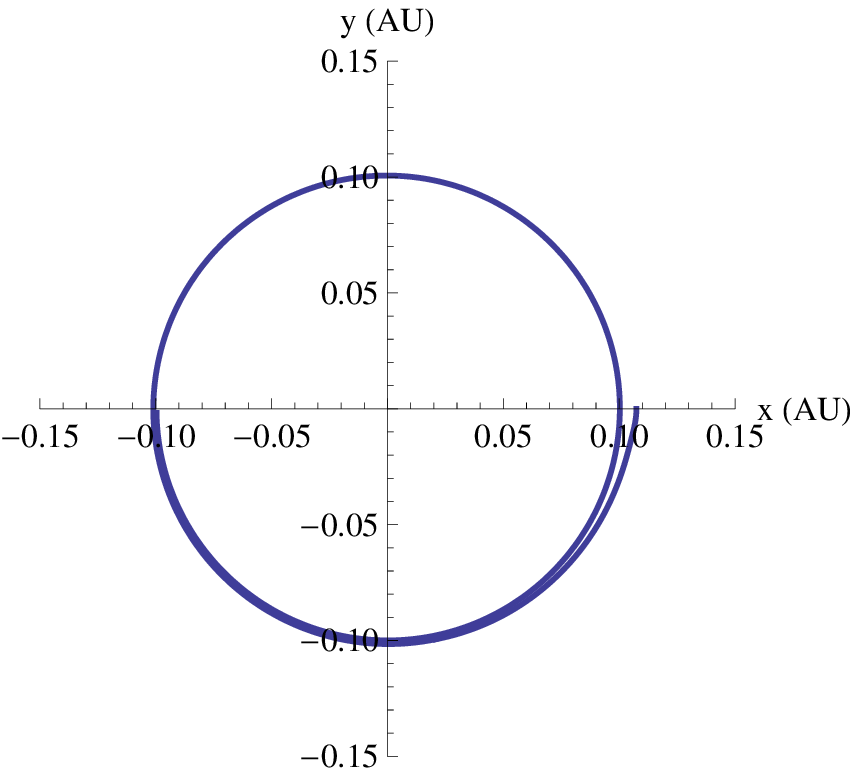}
\caption{Trajectory of charged grains projected in the midplane (in the reference framework corotating with the central star), in
which case X-ray flare occurs at time $t=0$ day and ends at time
$t=1$ day. }\label{fig:xyt}
\end{center}
\end{figure}

\begin{figure}[htbp]
\begin{center}
\includegraphics[width=14cm,height=8cm]{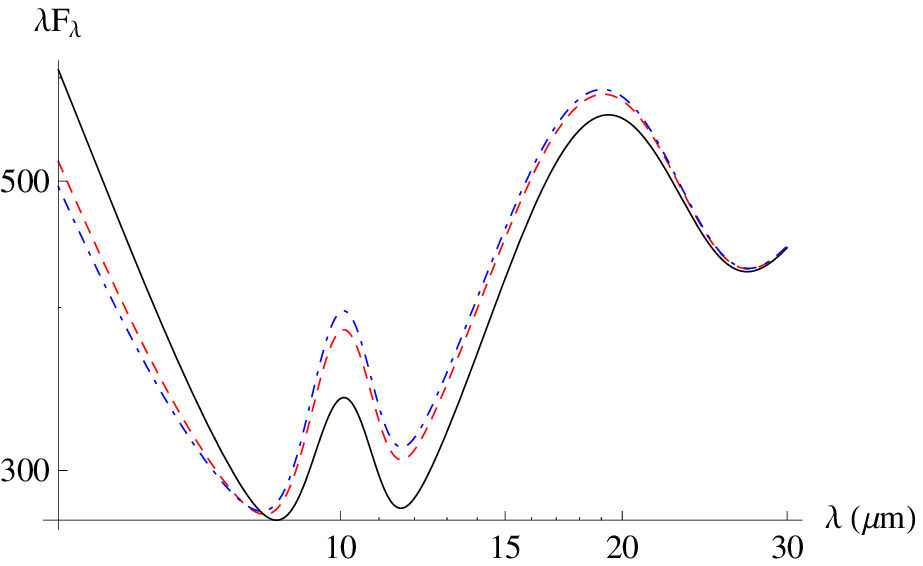}
\caption{Log-log plot of multiple spectra. The solid black line
represents the total spectrum of the star-disk system at time $t=0$
day; the dashed red line represents the total spectrum at time $t=1$
day, one day after the occurrence of X-ray flare; and the dot-dashed
blue line stands for the spectrum  at time $t=7$ day, six days after
the cease of X-ray flare. A clear seesaw feature is shown here: the
spectrum at shorter wavelength decreases and spectrum at longer
wavelength increases over time, with a pivot around $8\nu m$. The
spectral variability is around $20\%$ within one week.
}\label{fig:spectrum}
\end{center}
\end{figure}

We first describe the whole physical processes of spectral
variability triggered by X-ray flares. We assume stellar X-ray flare
with luminosity $\sim 10^{25} J/s$ starts at time $t=0$, and lasts
for one day. During X-ray outburst, gas in the inner rim gets
ionized quickly, with the ionization fraction rising from around
$4\%$ to around $80\%$ within $10^2$ seconds. Meanwhile the
X-ray photons lead to a temperture enhancement of the gas
in the disk surface layer from around $400K$ to
around $10^4K$ within the same timescale. As a result, speed of
electrons become higher, and the collisions between electrons and
grains become more frequently. In the equilibrium state, which can
be reached in almost no time, the charge on grains increases by
around $20$ times. Grains strongly negatively charged will be driven
by the magnetic field to undergo collective migration inward, in the
case that the moment of magnetic dipole is parallel with the disk
angular momentum. Consequently, the inner rim of the disk gets
squeezed, and the shadow casted by the grains on the flaring disk
attenuates. As a result, the spectrum at shorter wavelength
increases while the spectrum at longer wavelength decreases, as
sketched in Fig. \ref{fig:sketch}. In our numerical example, the
height of inner rim decreases by $17.6\%$ in the first day, and
decreases by $22.7\%$ cumulatively in seven days after occurrence
X-ray flares, as shown in Fig. \ref{fig:Ht}. In fact as discussed
before, location of the boundary of the shadowed region depends on
the ratio of height over radial distance of the inner rim, we show
variability of the height-radius ratio in Fig. \ref{fig:HRt}. In
Fig. \ref{fig:xyt}, we plot the trajectory of charged grains
projected in the midplane. We find newly charged grains still
undergo circular orbit, with a minor inward shift. Finally in Fig.
\ref{fig:spectrum}, we show the variability reproduced by our model.
A clear "seesaw" feature is shown here: the spectrum at shorter
wavelength decreases and spectrum at longer wavelength increases
over time, with a pivot around $8\mu m$. Spectrum at $\lambda=5\mu
m$ decreases by $14.9\%$ and $18.7\%$ respectively within one day
and the whole week; while spectrum at $\lambda=15\mu m$ increases by
$7.4\%$ and $9.2\%$ respectively within one day and the whole week.
The result is consistent with observations described in the first
section.

\section{Summary and Discussion}
In this section, we briefly discuss the influence of the direction
and strength of stellar magnetic dipole, as well as the grain size
on variability of the spectrum first, and then conclude the paper.

\begin{figure}[htbp]
\begin{center}
\includegraphics[width=14cm,height=8cm]{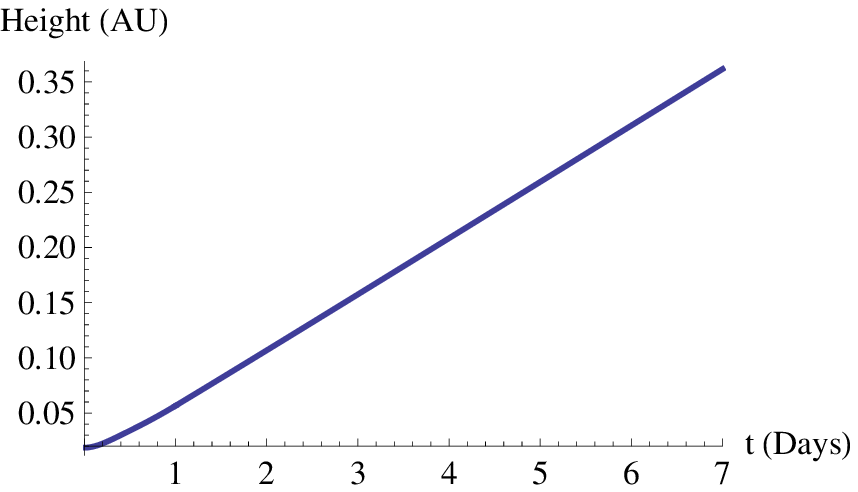}
\caption{In the anti-parallel case, height of inner rim increases
over time, in which case X-ray flare occurs at time $t=0$ day and
ends at time $t=1$ day.}\label{fig:HtA}
\end{center}
\end{figure}

\begin{figure}[htbp]
\begin{center}
\includegraphics[width=11cm,height=10cm]{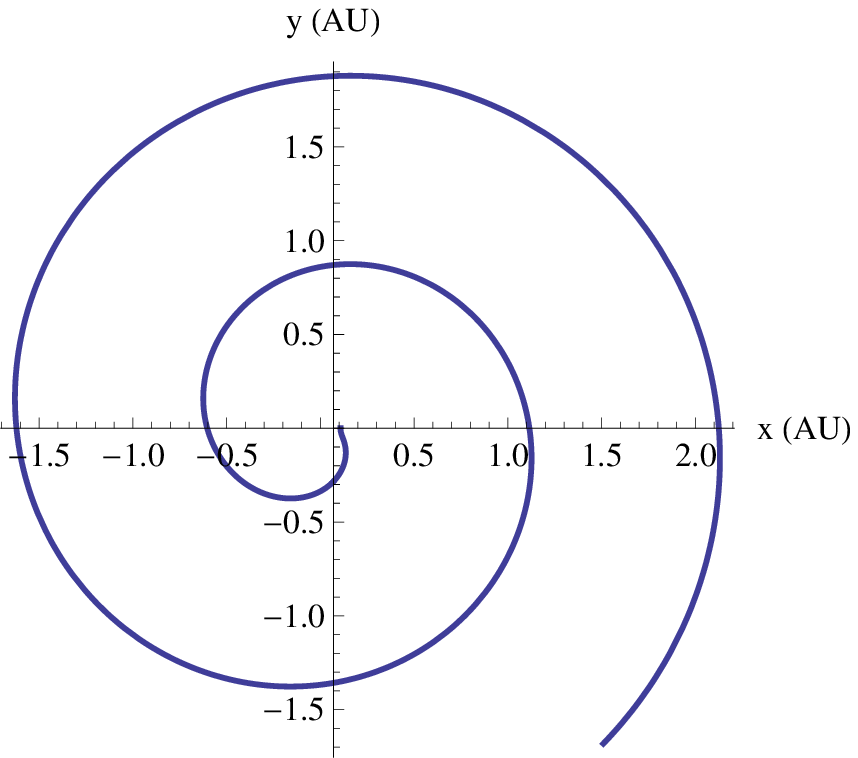}
\caption{Trajectory of charged grains projected in the midplane in
the anti-parallel case (in the reference framework corotating with the central star), in which case X-ray flare occurs at time
$t=0$ day and ends at time $t=1$ day.}\label{fig:xytA}
\end{center}
\end{figure}

In our numerical example above, the stellar magnetic moment is
parallel with the angular momentum of the disk. However, in a
general situation, the angle between the stellar magnetic moment and
angular momentum of the disk can range from $0$ to $\pi$, as
discussed by \citep{fla10}. Here we simply show the result for
anti-parallel case (with the angle as $\pi$). Contrary to the
parallel case, grains strongly negatively charged will be driven by
the magnetic field to undergo collective migration outward, so the
inner rim of the disk gets puffed up. The variability of the
shadowed region can be complicated: on one hand, the puffed-up inner
rim cast larger shadow on the flaring disk; on the other hand, some
stellar light may penetrate the attenuated puffed-up inner rim. As a
result, the spectrum at shorter wavelength increases while the
spectrum at longer wavelength may experience decreasing first and
increasing later. The variability of the height of inner rim is
shown in Fig. \ref{fig:HtA}, and the trajectory of charged grains
projected in the midplane is shown in Fig. \ref{fig:xytA}. Because
the gas gets much dilute outward, charged grains basically spiral
out of the system.

To look at the influence of magnetic field strength on variability
of the spectrum, we perform the same numerical simulation except for
stellar magnetic field with $B_s=500Gs$ and $B_s=2000Gs$
respectively in the parallel case. We expect stronger magnetic field
means stronger Lorentz force, resulting in more variability. In
fact, for stellar magnetic field with $B_s=500Gs$, the height of
inner rim decreases by $15.5\%$ in the first day, and decreases by
$20.6\%$ cumulatively in seven days after occurrence X-ray flares.
Spectrum at $\lambda=5\mu m$ decreases by $13.4\%$ and $17.2\%$
respectively within one day and the whole week; while spectrum at
$\lambda=15\mu m$ increases by $6.5\%$ and $8.5\%$ respectively
within one day and the whole week. The numerical result indicates
that weaker magnetic field can generate significant variability as
well. On contrary for stellar magnetic field with $B_s=2000Gs$, the
height of inner rim decreases by $19.7\%$ in the first day, and
decreases by $24.9\%$ cumulatively in seven days after occurrence
X-ray flares. Spectrum at $\lambda=5\mu m$ decreases by $16.4\%$ and
$20.1\%$ respectively within one day and the whole week; while
spectrum at $\lambda=15\mu m$ increases by $8.2\%$ and $9.9\%$
respectively within one day and the whole week. Spectrum indeed
becomes more volatile for stronger stellar magnetic field.

To look at the influence of grain size on variability of the
spectrum, we perform the same numerical simulation except that grain
radius $r_d=0.1\mu m$ and $r_d=1\mu m$ respectively in the parallel
case. Mass of grains is proportional to $r_d^3$, while the Lorentz
force is proportional to $Q_0$, which is proportional to $r_d$,
thus smaller grains experience larger force per unit mass, therefore
are expected to display more variability. In fact, for grains with
radius $r_d=0.1\mu m$, the height of inner rim decreases by $22.5\%$
in the first day, and decreases by $21.7\%$ cumulatively in seven
days after occurrence X-ray flares. Spectrum at $\lambda=5\mu m$
decreases by $21.7\%$ and $26.7\%$ respectively within one day and
the whole week; while spectrum at $\lambda=15\mu m$ remains almost
unchanged with decrease by $0.5\%$ and $1\%$ respectively within one
day and the whole week. The numerical result is consistent with our
analysis, and the spectrum also decreases at longer wavelength due
to a sharp decrease in the spectrum from the inner rim. On contrary
for grains with radius $r_d=1\mu m$, the height of inner rim
decreases by $15.5\%$ in the first day, and decreases by $20.8\%$
cumulatively in seven days after occurrence X-ray flares. Spectrum
at $\lambda=5\mu m$ decreases by $11.9\%$ and $14.8\%$ respectively
within one day and the whole week; while spectrum at $\lambda=15\mu
m$ increases by $8.1\%$ and $10.7\%$ respectively within one day and
the whole week.

In conclusion, we construct a physical model to attribute the
puzzling rapid MIR spectral variability to the observed X-ray flares
from T Tauri stars. Our simulation result captures all main observed
features consistently, and is robust with respect to different grain
size and strength of stellar magnetic field. We made various
assumptions in the construction of this model, including the
location of the $R_{rim}$ and an assumed distribution of the disk
aspect ratio.  The latter can be mapped out with regular monitor of
disk's SED changes.  Both of these quantities may also depend on the
stellar magnetic field.  It may be useful to explore correlations,
if any, between these observable quantities. Finally, the basic
assumption that the SED variations are triggered by X-ray flares may
be tested by simultaneous X-ray and infrared observations.






\acknowledgments
\bf{Acknowledgments}

\rm

We thank Xu Huang and Wenhua Ju for their constructive suggestions.
This work is supported by NASA
(NNX07A-L13G, NNX07AI88G, NNX08AL41G, NNX08AM84G), and NSF(AST-0908807).

\end{document}